\documentclass[useAMS,usenatbib]{mn2e}

\usepackage{graphicx}
\usepackage{amssymb}

\newcommand{\apj}{ApJ}

\newcommand{\aap}{A\&A}

\newcommand{\aj}{AJ}
\newcommand{\mnras}{MNRAS}

\newcommand{\apjs}{ApJS}
\newcommand{\MC}{\multicolumn}
\newcommand{\kms}{km~s$^{-1}$}
\newcommand{\Te}{T$_{\rm e}$}
\newcommand{\Hb}{H$\beta$}
\newcommand{\p}{$\pm$}
\newcommand{\sunn}{$_{\odot}$}
\DeclareRobustCommand{\ion}[2]{%
\relax\ifmmode
\ifx\testbx\f
{\mathrm{#1\,\textsc{#2}}}\else
{\mathrm{#1\,\mathsc{#2}}}\fi
\else\textup{#1\,{\mdseries\textsc{#2}}}%
\fi}

\title[SDSS J092609.45+334304.1: a nearby unevolved galaxy]
{SDSS J092609.45+334304.1: a nearby unevolved galaxy}
\author[S.A. Pustilnik, A.L. Tepliakova, A.Y. Kniazev, J.-M. Martin,
A.N. Burenkov]
{S.A. Pustilnik,$^{1,5}$\thanks{E-mail: sap@sao.ru (SAP)}
A.L. Tepliakova,$^1$
A.Y. Kniazev,$^{2,3}$
J.-M. Martin,$^{4}$
and A.N. Burenkov$^1$\\
\rule{-4pt}{20pt}
$^1$ Special Astrophysical Observatory of RAS, Nizhnij Arkhyz,
  Karachai-Circassia 369167, Russia\\
$^2$ South African Astronomical Observatory, PO Box 9, 7935 Observatory,
   Cape Town, South Africa\\
$^3$ Southern African Large Telescope Foundation, PO Box 9, 7935 Observatory,
   Cape Town, South Africa\\
$^4$ GEPI and Station de radioastronomie, Observatoire de Paris, 5 place Jules Janssen, 92190 Meudon, France  \\
$^5$ Isaac Newton Institute of Chile, SAO branch, Nizhnij Arkhyz, Russia}

\begin{document}

\label{firstpage}

\date{Accepted 2009 August 30. Received 2009 July 3}

\pagerange{\pageref{firstpage}--\pageref{lastpage}} \pubyear{2009}

\maketitle

\begin{abstract}

We present the results of observations of the very low surface brightness
(VLSB) dwarf galaxy SDSS J092609.45+334304.1  with extreme parameters
which indicate its unevolved status. Namely, its value of O/H, derived as an
average of that in two adjacent \ion{H}{ii} regions at the NE edge of the
disc, corresponds to the parameter 12+$\log$(O/H)=7.12$\pm$0.02, which is
amongst two lowest known.
The total
HI flux measurement obtained with the
Nan\c {c}ay Radio Telescope and the photometric results imply that the galaxy
ratio M(HI)/L$_{\rm B} \sim$3.0, is among the top known in the Local
Volume. The galaxy is situated in the region of a nearby underdense region
known as the Lynx-Cancer void, where other, unevolved galaxies, including
DDO~68, HS~0832+3542 and SAO~0822+3545, are  known to be present.
The total mass of this almost edge-on VLSB galaxy is $\sim$8.3 times larger
than its baryonic mass, implying the dynamical dominance of Dark Matter (DM)
halo. The $(u-g), (g-r)$ colours of outer parts of this galaxy are
consistent with the ages of its main stellar population of 1--3 Gyr.
Thanks to the galaxy isolation, the small effect of current or recent
star formation (SF), its proximity and rather large HI flux
($\sim$2.5~Jy$\cdot$\kms), this VLSB dwarf
is a good laboratory for the detailed study of DM halo properties
through HI kinematics and the star formation processes in very metal-poor
low surface density environment. This finding, along with the discovery
of other unusual dwarf galaxies in this void,  provides evidence for
the relation between galaxy evolution and its very low-density environment
for the baryonic mass range of 10$^{8}$ to 10$^{9}$ M\sunn.
This relation seems to be consistent with that expected in
the $\Lambda$CDM models of galaxy and structure formation.

\end{abstract}


\section[]{INTRODUCTION}
\label{sec:intro}

The modern cosmological CDM models of the large-scale structure and galaxy
formation, including the state-of-art N-body simulations, predict that
galaxy properties and evolution can significantly depend on their global
environment
\citep[e.g.,][ and references therein]{Peebles01, MW02, Gottlober03, Hoeft06,
Arkhipova07, Hahn07, Hahn09}.
While the effect of a denser environment on galaxy properties and evolution
is known for rather long time \citep[e.g.,][]{HGC84},
the role of the most rarefied environment (typical of
voids) on galaxy formation and evolution is less studied both theoretically
and observationally. For observational aspects, various selection effects
must be taken into account.
In particular, most galaxies with known radial velocities come
from spectral surveys of magnitude-limited samples. Wide-field
spectral surveys have typical magnitude limits corresponding roughly to
$B \sim$18.
This apparent magnitude limit implies that for distances well beyond the
Local Supercluster ($cz > 5000-6000$~\kms), where large voids were found,
the faintest selected galaxies will have the absolute magnitudes M$_{\rm B}$
of $\sim$--16 or higher. This implies that the study of distant voids is
limited by the galaxies with M$_{\rm B}$ of only $\sim$2--3 mag. fainter
than that corresponding to the luminosity of L$_{*}$ (M$_{\rm B}^{*}
\sim$--19.5). The latter (or close to this) is usually used for selection of
galaxies delineating the borders of `empty' regions - voids.
Therefore, even the most advanced studies of void galaxy population, based
on very large samples with redshifts from the Sloan Digital Sky Survey (SDSS)
\citep[e.g.,][]{Sorrentino06,Patiri06} of $z <$0.03--0.05, were limited by
the galaxy samples with
luminosities only 1.5-2~magnitudes fainter than M$_{\rm B}^{*} \sim$--19.5.
Since the possible difference of galaxy properties in various types of
environment is expected to depend on galaxy mass, these `shallow' probes
of `distant' void galaxy population, based on the SDSS samples,
may be inadequate to get  deeper insights into the questions.

In order to probe the properties of smaller luminosities void galaxies,
a wide-angle redshift survey with a similar magnitude limit is needed,
for a volume located several times closer than those used in
the previous studies.
Apart from the giant Local Void described by \citet{Tully08}, which
extends to tens Mpc and appears almost empty, several relatively small
voids adjacent to
the Local Volume \citep[defined by D $<$ 10~Mpc, e.g.,][]{CNG} are identified
\citep[see][]{Fairall98}. One more such void in Lynx-Cancer sky region was
noticed by \citet{SAO0822}. Due to its relative proximity
(D$_{\rm centre}$ $\sim$14~Mpc), galaxies selected in this region for
the SDSS spectroscopy, have the absolute magnitudes down to M$_{\rm B}$ of
--(12.5-13.0). Studying the dwarf galaxies population in nearby voids, and
in particular, in this Lynx-Cancer void, will permit to advance significantly
in probing void dwarf galaxy properties. This will also give a base for
a comparison of the {\it evolutionary status} of void dwarf galaxies and
those located in the regions with the higher galaxy density.

Coming back to the study of void galaxy population, it is worth noting the
well known correlation between galaxy luminosities and their central surface
brightnesses  \citep[e.g.,][ and references therein]{Cross02}.
Since most of voids galaxies are dwarfs (as seen, in particular, in the
sample of galaxies falling on Lynx-Cancer void), with the major fraction
by number  of low luminosities, the proportion of LSB galaxies in voids
is expected to be  also higher. In the SDSS project, relatively high
 surface brightness (SB) lower limits have been used for spectroscopic
target selection.
Therefore, the completeness for the SDSS redshift samples falls below 50\% for
objects with the half-light $\mu_{50,r} \geq$ 23.5~mag~arcsec$^{-2}$
\citep[e.g.,][]{Geha06}. Hence, the resulting SDSS galaxy samples with
redshifts are biased against LSB galaxies and this can lead to a substantial
loss of LSB dwarf galaxies in void galaxy samples.
Some of these missing galaxies might form the youngest local galaxy population
\citep{Zackrisson05}.

Blind HI surveys can overcome this selection effect, at least for late-type
galaxies.
Therefore, one can expect that the Arecibo blind HI survey ALFALFA
\citep[e.g.,][]{ALFALFA} will further increase the current Lynx-Cancer void
dwarf sample and will find mainly LSBD new galaxies. The currently available
dwarf sample includes about fifty late-type galaxies (Pustilnik et
al., in preparation).

During the systematic study of dwarf galaxies in the Lynx-Cancer void we have
already
discovered several unusual objects, including DDO~68
\citep*{DDO68,IT07,DDO68_sdss}, HS~0822+3542, SAO~0822+3545
\citep{HS0822,SAO0822,Chengalur06}.
In this paper we report on the case study of one of the Lynx-Cancer void
genuine LSB dwarfs, which passed through the SDSS spectral target SB selection
criterion because of its almost edge-on viewing direction and its
brightening in apparent SB. It appeared as an extremely metal-poor and
very gas-rich galaxy located  remarkably close to DDO~68.
The paper is organised as follows. In Section~\ref{sec:obs} we briefly
describe
the observations and the reduction of obtained data. Section
~\ref{sec:results} presents the results of observations and their analysis.
In Section~\ref{sec:dis} we discuss the results and their implications
in a  broader context and summarise our conclusions.

\section[]{OBSERVATIONS AND DATA REDUCTION}
\label{sec:obs}

\subsection{Optical observations}

\begin{table*}
\begin{center}
\caption{Journal of the 6\,m telescope observations of SDSS~J0926+3343}
\label{Tab1}
\begin{tabular}{lrccccccc} \\ \hline \hline
\MC{1}{c}{ Date }       &
\MC{1}{c}{ Exposure }   &
\MC{1}{c}{ Wavelen.Range [\AA] } &
\MC{1}{c}{ Dispersion } &
\MC{1}{c}{ Spec.resol. } &
\MC{1}{c}{ Seeing }     &
\MC{1}{c}{ Airmass }     &
\MC{1}{c}{ Grism }       &
\MC{1}{c}{ Detector }    \\

\MC{1}{c}{ }       &
\MC{1}{c}{ time [s] }    &
\MC{1}{c}{           } &
\MC{1}{c}{ [\AA/pixel] } &
\MC{1}{c}{ FWHM(\AA) } &
\MC{1}{c}{ [arcsec] }    &
\MC{1}{c}{          }    &
\MC{1}{c}{          }    &
\MC{1}{c}{          }     \\

\MC{1}{c}{ (1) } &
\MC{1}{c}{ (2) } &
\MC{1}{c}{ (3) } &
\MC{1}{c}{ (4) } &
\MC{1}{c}{ (5) } &
\MC{1}{c}{ (6) } &
\MC{1}{c}{ (7) } &
\MC{1}{c}{ (8) } &
\MC{1}{c}{ (9) } \\
\hline
\\[-0.3cm]
 2008.11.26  & 4$\times$900 & $ 3500-7500$ & 2.1 & 12.0 & 2.3 & 1.11 & VPHG550G  & 2K$\times$2K  \\
 2009.01.21  & 6$\times$900 & $ 3700-5700$ & 0.9 & 5.5  & 1.3 & 1.11 & VPHG1200G & 2K$\times$4K  \\
 2009.01.22  & 4$\times$900 & $ 3600-9000$ & 2.1 & 12.0 & 1.5 & 1.02 & VPHG550G  & 2K$\times$4K  \\
 2009.02.19  & 8$\times$900 & $ 3500-7500$ & 2.1 & 12.0 & 1.9 & 1.05 & VPHG550G  & 2K$\times$2K  \\
\hline \hline \\[-0.2cm]
\end{tabular}
\end{center}
\end{table*}

The long-slit spectral observations of galaxy SDSS J092609.45+334304.1
(hereafter J0926+3343 for brevity, see its main parameters in
Table~\ref{tab:param})
were conducted with the multimode instrument SCORPIO \citep{SCORPIO}
installed at the prime focus of the SAO 6\,m telescope (BTA) on the
nights of 2008 November 26, 2009 January 21 and 22, and 2009 February 19.
The grism VPHG550G was used with the 2K$\times$2K CCD detector EEV~42-40
on 2008 November 26 and 2009 February 19. On 2009 January 21 and 22, the
grisms VPHG1200G and VPHG550G with the 2K$\times$4K CCD detector EEV~42-90
were used.
The respective spectral ranges, spectral resolution, exposure times and
seeings are presented in Table~\ref{Tab1}.
The scale along the slit (after binning) was 0\farcs36 pixel$^{-1}$ in all
cases. The object spectra were complemented before or after by the reference
spectra of He--Ne--Ar lamp for the wavelength calibration. The spectral
standard star Feige~34 \citep{Bohlin96} was observed during the nights for
the flux calibration.

\begin{figure}
 \centering
 \includegraphics[angle=-0,width=6.8cm]{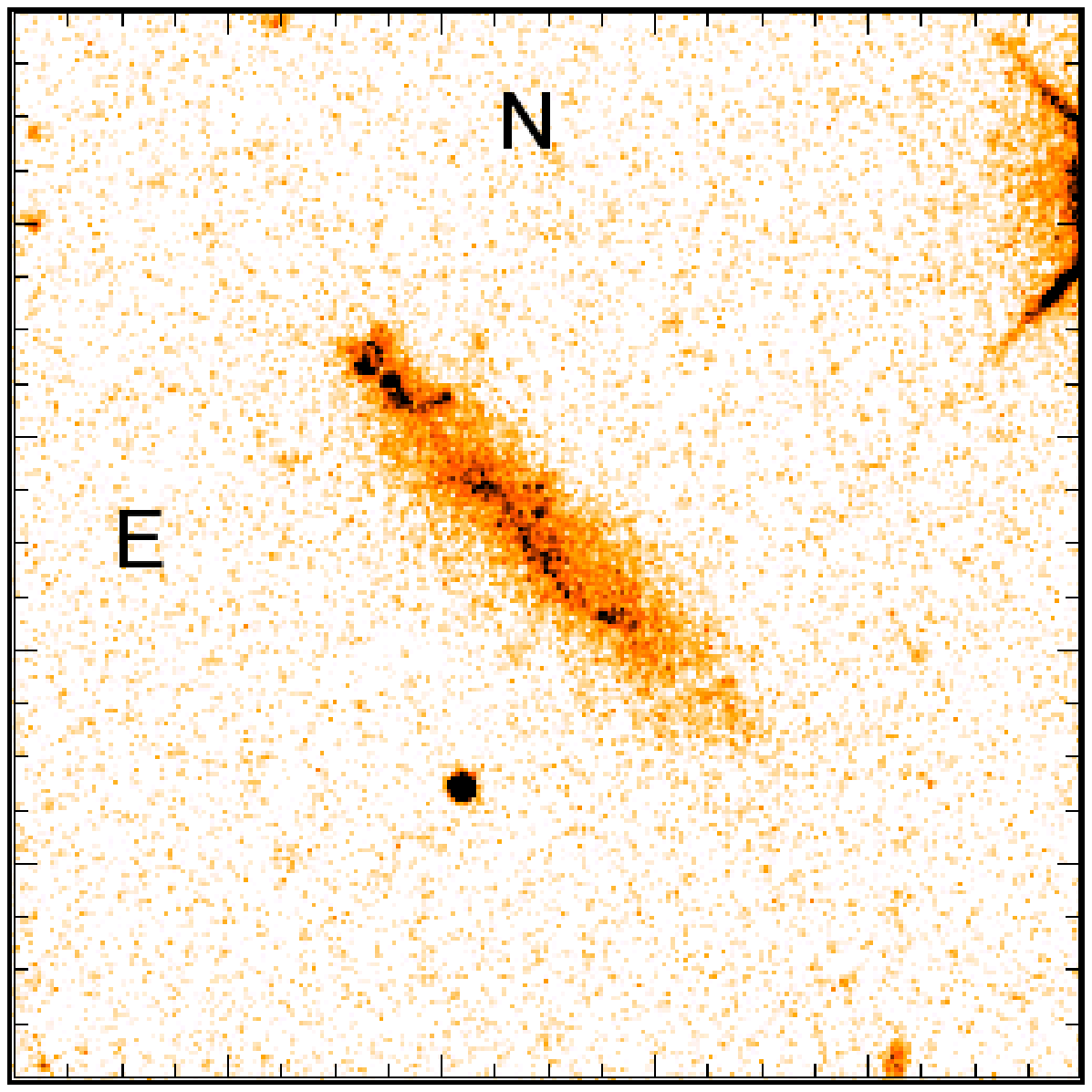}
 \includegraphics[angle=-0,width=6.8cm]{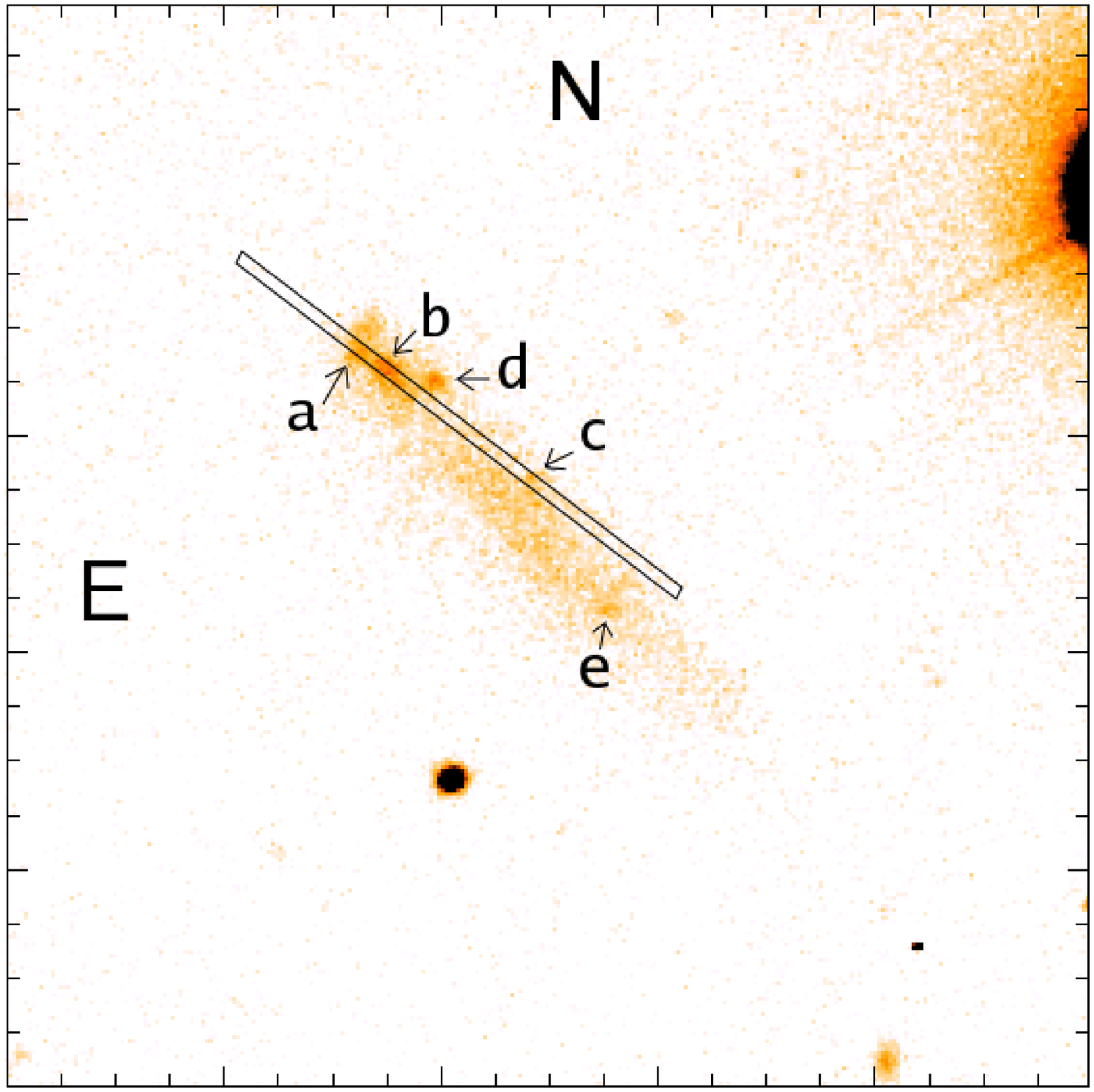}
 \includegraphics[angle=-0,width=6.8cm, clip=]{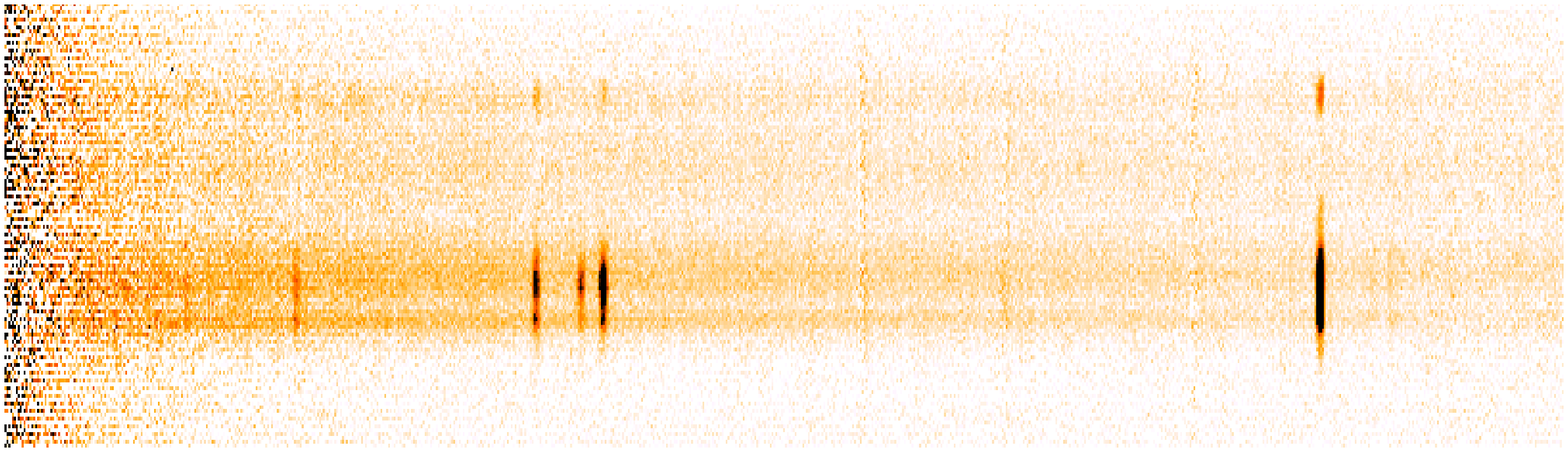}
  \caption{\label{fig:image}
{\bf Top panel}:
The $g$-filter image of SDSS J0926+3343 obtained from the
SDSS database. The total size of the field is 100\arcsec\ by 100\arcsec.
Some `knots' at the NE edge are faint \ion{H}{ii} regions. The brightness
enhancement seen as a ridge near the LSBD centre along the major axis looks
like quite curved and patchy, indicating relatively recent SF episode.
{\bf Middle panel:}
Two faint  \ion{H}{ii} regions {\bf a}, {\bf b} ($\sim$4\arcsec\ in between)
at the NE edge of galaxy disc and  region {\bf c} closer to the galaxy
centre (indicated by arrows) are shown in the BTA H$\alpha$-filter (SED665)
image with the position of the long slit overlaid (PA=54\degr).
The regions  {\bf d}, mentioned in Sect.~\ref{sect:age}, and the faint
\ion{H}{ii}-region {\bf e}, for which the SDSS spectrum was obtained,
are also indicated.
The size of the field is 100\arcsec\ by 100\arcsec.
In the {\bf bottom panel} the BTA 2D spectrum of SDSS J0926+3343 is shown
with the emission of \ion{H}{ii} regions {\bf a}, {\bf b} in the lower part, and
the faint emission  of the \ion{H}{ii} region {\bf c} seen in the upper part
of spectrum.
}
\end{figure}

The long slit was positioned on the two \ion{H}{ii} regions {\bf a} and
{\bf b}
on the NE edge of SDSS J0926+3343 edge-on disc (see images of the galaxy in
the SDSS $g$-filter and BTA filter SED665 in Fig.~\ref{fig:image}). These
\ion{H}{ii} regions (with a distance of $\sim$4\arcsec\ in between) were
identified through one-minute exposure acquisition images of this galaxy
with the medium-width filter SED665, centred at $\lambda$6622~\AA\ and with
FWHM=191~\AA. The slit was centred on these \ion{H}{ii} regions. Its
direction was not along the disc major axis. The spectrum of one more,
a fainter \ion{H}{ii} region {\bf c} at $\sim$17\arcsec\
from {\bf b}, close to the galaxy centre, was acquired at this slit
orientation. We notice that the region {\bf c} is different from the
very faint region, labeled in Fig.~\ref{fig:image} as {\bf e},
for which the SDSS spectrum was obtained.
The latter is situated at $\sim$31\arcsec\ from region {\bf b}.

All spectral data reduction and emission line
measurements were performed similar to that described in \citet{DDO68}.
Namely, the standard pipeline with the use of IRAF\footnote{IRAF: the Image
Reduction and Analysis Facility is distributed by the National Optical
Astronomy Observatory, which is operated by the Association of Universities
for Research in Astronomy, Inc. (AURA) under cooperative agreement with the
National Science Foundation (NSF).}
and {\tt MIDAS}\footnote{MIDAS is an acronym for the European Southern
Observatory package -- Munich Image Data Analysis System. }
was applied for the reduction of long-slit spectra, which included the
following steps: removal of cosmic ray hits,
bias subtraction,  flat-field correction, wavelength
calibration, night-sky background subtraction. Then, using the data on the
spectrophotometry standard star, all spectra were transformed to absolute
fluxes. We could extract individual 1D spectra of regions {\bf a} and {\bf b}
only for nights 2009.01.21 and 2009.01.22. This was done by summing up
without weighting of 6 and 10 rows along the slit, respectively.

In order to derive the estimate of the galaxy O/H using all obtained data,
we extracted also for every night the light of the `composite' spectrum,
containing both regions {\bf a} and {\bf b}, summing up without weighting
21 rows ($\sim$7.5\arcsec)
along the slit (see more details in Section~\ref{sec:results}).
The emission line intensities with their errors were measured in the way
described in detail in \citet{SHOC}.

\subsection{HI observations}

The \ion{H}{i}-observations with the
Nan\c {c}ay\footnote{The Nan\c {c}ay Radioastronomy Station is part of the
Observatoire de Paris and is operated by the Minist\`ere de l'Education
Nationale and Institut des Sciences de l'Univers of the Centre National
de la Recherche Scientifique.}
radio telescope (NRT) with a collecting area of
200$\times$34.5~m are characterised by a half-power beam width (HPBW) of
3.7\arcmin~(East-West)$\times$22\arcmin~(North-South) at
declination $\delta$=0\degr\
(see also \verb|http://www.obs-nancay.fr/nrt|).
The data were acquired during July-September 2007, with the total time
on-source of $\sim$6 hours. We used the antenna/receiver system F.O.R.T.
\citep[e.g.,][]{FORT} with improved overall sensitivity.
The system temperature was $\sim$35 K for both the horizontal and vertical
linear polarisations of a dual-polarisation receiver.
The gain of the telescope was 1.5 K~Jy$^{-1}$ at declination $\delta$=0\degr.
The 8192-channel correlator was used covering a total bandwidth of 12.5 MHz.
The total velocity range covered
was about 2700~\kms, with the channel spacing of 1.3~\kms\ before smoothing.
The observations consisted of separate cycles of `ON' and `OFF' integrations,
each of 40 seconds in duration. `OFF' integrations were acquired at the
target declination, with the East R.A. offset of
$\sim$15\arcmin~$\times$ cos($\delta$).
For more detail see the description in the paper by \citet{NRT_07}.

The data were reduced using the NRT standard programs NAPS and SIR,
written by the telescope staff (see description on
\mbox{http://www.obs-nancay.fr/nrt/support}).
Horizontal and vertical polarisation spectra were
calibrated and processed independently and then averaged
together. The error estimates were calculated following
\citet{Schneider86}.
The baselines were generally well-fit by a third order or lower
polynomial  and were subtracted out.

\subsection{Imaging data from the SDSS}

The SDSS \citep{York2000} is well suited for
photometric studies of various galaxy samples due to its homogeneity, area
coverage, and depth (SDSS Project Book\footnote{
http://www.astro.princeton.edu/PBOOK/science/\\galaxies/galaxies.htm}).
SDSS is an imaging and spectroscopic survey that covers about
one-quarter of the Celestial Sphere. The imaging data are collected in drift
scan mode in five bandpasses \citep[$u, \ g, \ r, \ i$, and $z$;][]{SDSS_phot}
using mosaic CCD camera \citep{Gunn98}. An automated image-processing system
detects astronomical sources and measures their photometric and astrometric
properties \citep{Lupton01,SDSS_phot1,Pier03} and identifies candidates for
multi-fibre spectroscopy.
At the same time, the pipeline-reduced SDSS data can be used, if
necessary, in order to get independent photometry \citep[e.g.,][]{Kniazev04}.
For our current study the images in the respective filters were retrieved from
the SDSS Data Release 7 \citep[DR7;][]{DR7}.

Since the SDSS provides users with the fully reduced images, the only
additional step we needed to perform (apart from the photometry in round
diaphragms) was the background subtraction. For this, all bright stars were
removed from the images. After that the studied object was masked and the
background level within this mask was approximated with the package {\tt aip}
from {\tt MIDAS}.  In more detail the method and the related programs are
described in \cite{Kniazev04}.
To transform instrumental fluxes in diaphragms to stellar magnitudes, we
used the photometric system coefficients defined in SDSS for the used field.
The accuracy of zero-point determination was  $\sim$0.01 mag in all
filters.

\section[]{RESULTS}
\label{sec:results}

\subsection[]{Spectra and oxygen abundance}

\begin{table*}
\centering{
\caption{Line intensities in `composite' spectra of \ion{H}{ii} regions {\bf a} and {\bf b}}
\label{tab:Intens1}
\begin{tabular}{lcccccccc} \hline
\rule{0pt}{10pt}
& \MC{2}{c}{2008.11.26} & \MC{2}{c}{2009.01.21} & \MC{2}{c}{2009.01.22} & \MC{2}{c}{2009.02.19} \\ \cline{2-3} \cline{4-5} \cline{6-7} \cline{8-9}
\rule{0pt}{10pt}
$\lambda_{0}$(\AA) Ion                    &
$F$/$F$(H$\beta$)&$I$/$I$(H$\beta$) &
$F$/$F$(H$\beta$)&$I$/$I$(H$\beta$) &
$F$/$F$(H$\beta$)&$I$/$I$(H$\beta$) &
$F$/$F$(H$\beta$)&$I$/$I$(H$\beta$) \\ \hline

3727\ [O\ {\sc ii}]\            &  55.5$\pm$8.9  &  52.8$\pm$8.9   &  24.5$\pm$29.4&  24.5$\pm$29.4&  43.3$\pm$9.8 &  43.6$\pm$10.9&  58.5$\pm$7.8 &  55.9$\pm$8.4   \\
4101\ H$\delta$\                &  20.1$\pm$2.8  &  26.4$\pm$4.5   &  25.1$\pm$3.4 &  25.1$\pm$3.4 &   9.7$\pm$1.3 &  24.4$\pm$4.4 &  12.0$\pm$1.5 &  28.4$\pm$5.0   \\
4340\ H$\gamma$\                &  39.3$\pm$3.8  &  42.9$\pm$4.7   &  48.3$\pm$3.8 &  48.3$\pm$3.8 &  38.1$\pm$2.2 &  48.4$\pm$3.4 &  34.7$\pm$1.6 &  46.8$\pm$2.9   \\
4363\ [O\ {\sc iii}]\           &   6.4$\pm$2.9  &   6.1$\pm$2.9   &   6.0$\pm$2.7 &   6.0$\pm$2.7 &   4.6$\pm$1.7 &   4.3$\pm$1.8 &   5.2$\pm$1.2 &   4.7$\pm$1.3   \\
4861\ H$\beta$\                 &   100$\pm$5.5  &   100$\pm$6.0   &   100$\pm$5.1 &   100$\pm$5.9 &   100$\pm$3.8 &   100$\pm$3.9 &   100$\pm$3.5 &   100$\pm$4.0   \\
4959\ [O\ {\sc iii}]\           &  57.4$\pm$4.0  &  55.0$\pm$4.0   &  61.1$\pm$4.0 &  61.1$\pm$4.0 &  64.0$\pm$3.0 &  57.7$\pm$3.1 &  57.9$\pm$2.4 &  51.3$\pm$2.4   \\
5007\ [O\ {\sc iii}]\           &   171$\pm$8.2  &   163$\pm$8.2   &   182$\pm$7.9 &   182$\pm$7.9 &   188$\pm$6.3 &   168$\pm$6.3 &   182$\pm$5.7 &   160$\pm$5.6   \\
5876\ He\ {\sc i}\              &  10.7$\pm$1.8  &  10.2$\pm$1.8   &   9.2$\pm$2.4 &   9.2$\pm$2.4 &   9.0$\pm$1.7 &   7.6$\pm$1.6 &   8.5$\pm$0.8 &   7.2$\pm$0.8   \\
6548\ [N\ {\sc ii}]\            &   1.7$\pm$2.0  &   1.7$\pm$2.0   &      ...      &      ...      &   0.3$\pm$1.5 &   0.2$\pm$1.3 &   1.0$\pm$0.8 &   0.9$\pm$0.7   \\
6563\ H$\alpha$\                &   283$\pm$12.6 &   272$\pm$13.8  &      ...      &      ...      &   335$\pm$10.4&   277$\pm$10.3&   327$\pm$9.0 &   275$\pm$9.3   \\
6584\ [N\ {\sc ii}]\            &   5.9$\pm$3.7  &   5.6$\pm$3.7   &      ...      &      ...      &   0.8$\pm$3.2 &   0.7$\pm$2.9 &   3.4$\pm$1.9 &   2.8$\pm$1.8   \\
& &  &  & \\
C(H$\beta$)\ dex                & \MC {2}{c}{0.00$\pm$0.06} & \MC {2}{c}{0.00$\pm$0.07} & \MC {2}{c}{0.14$\pm$0.04} & \MC {2}{c}{0.10$\pm$0.04}  \\
EW(abs)\ \AA\                   & \MC {2}{c}{2.75$\pm$0.72} & \MC {2}{c}{0.00$\pm$2.38} & \MC {2}{c}{5.25$\pm$0.45} & \MC {2}{c}{5.90$\pm$0.43}  \\
$F$(H$\beta$)$^a$\              & \MC {2}{c}{5.12$\pm$0.20} & \MC {2}{c}{6.33$\pm$0.23} & \MC {2}{c}{7.31$\pm$0.19} & \MC {2}{c}{5.14$\pm$0.15}  \\
EW(H$\beta$)\ \AA\              & \MC {2}{c}{ 55.2$\pm$2.1} & \MC {2}{c}{ 81.2$\pm$3.0} & \MC {2}{c}{ 52.0$\pm$1.4} & \MC {2}{c}{ 48.0$\pm$1.2}  \\
V$_{hel}$\ \kms\                & \MC {2}{c}{ 567$\pm$66}   & \MC {2}{c}{ 558$\pm$12}   & \MC {2}{c}{ 465$\pm$30}   & \MC {2}{c}{ 492$\pm$39}    \\ \hline
\MC{5}{l}{$^a$ in units of 10$^{-16}$ ergs\ s$^{-1}$cm$^{-2}$.}
\end{tabular}
 }
\end{table*}

\begin{table*}
\centering{
\caption{Abundances derived from `composite' spectra of regions {\bf a} and {\bf b}}
\label{t:Chem1}
\begin{tabular}{lcccc} \hline
\rule{0pt}{10pt}
\rule{0pt}{10pt}
Value                                      & 2008.11.26      & 2009.01.21      & 2009.01.22      & 2009.02.19        \\   \hline
$T_{\rm e}$(OIII)(10$^{3}$~K)\             & 21.46$\pm$6.72  & 19.72$\pm$5.32  & 17.12$\pm$3.73  & 18.66$\pm$2.84    \\
$T_{\rm e}$(OII)(10$^{3}$~K)\              & 16.08$\pm$1.62  & 15.58$\pm$5.34  & 14.71$\pm$3.84  & 15.42$\pm$2.90    \\
$N_{\rm e}$(SII)(cm$^{-3}$)\               &   10$\pm$10~~   &  10$\pm$10~~    &   10$\pm$10~~   &  10$\pm$10~~      \\
& & &  &\\
O$^{+}$/H$^{+}$($\times$10$^{-5}$)\        & 0.389$\pm$0.132 & 0.198$\pm$0.312 & 0.420$\pm$0.360 & 0.465$\pm$0.274   \\
O$^{++}$/H$^{+}$($\times$10$^{-5}$)\       & 0.808$\pm$0.498 & 1.075$\pm$0.602 & 1.360$\pm$0.679 & 1.049$\pm$0.345   \\
O/H($\times$10$^{-5}$)\                    & 1.197$\pm$0.515 & 1.273$\pm$0.678 & 1.780$\pm$0.768 & 1.515$\pm$0.441   \\
12+log(O/H)\                               & ~7.08$\pm$0.19~ & ~7.10$\pm$0.23~ & ~7.25$\pm$0.19~ & ~7.18$\pm$0.13~   \\
\hline
\end{tabular}
 }
\end{table*}

\begin{figure}
  \centering
 \includegraphics[angle=-90,width=7.5cm, clip=]{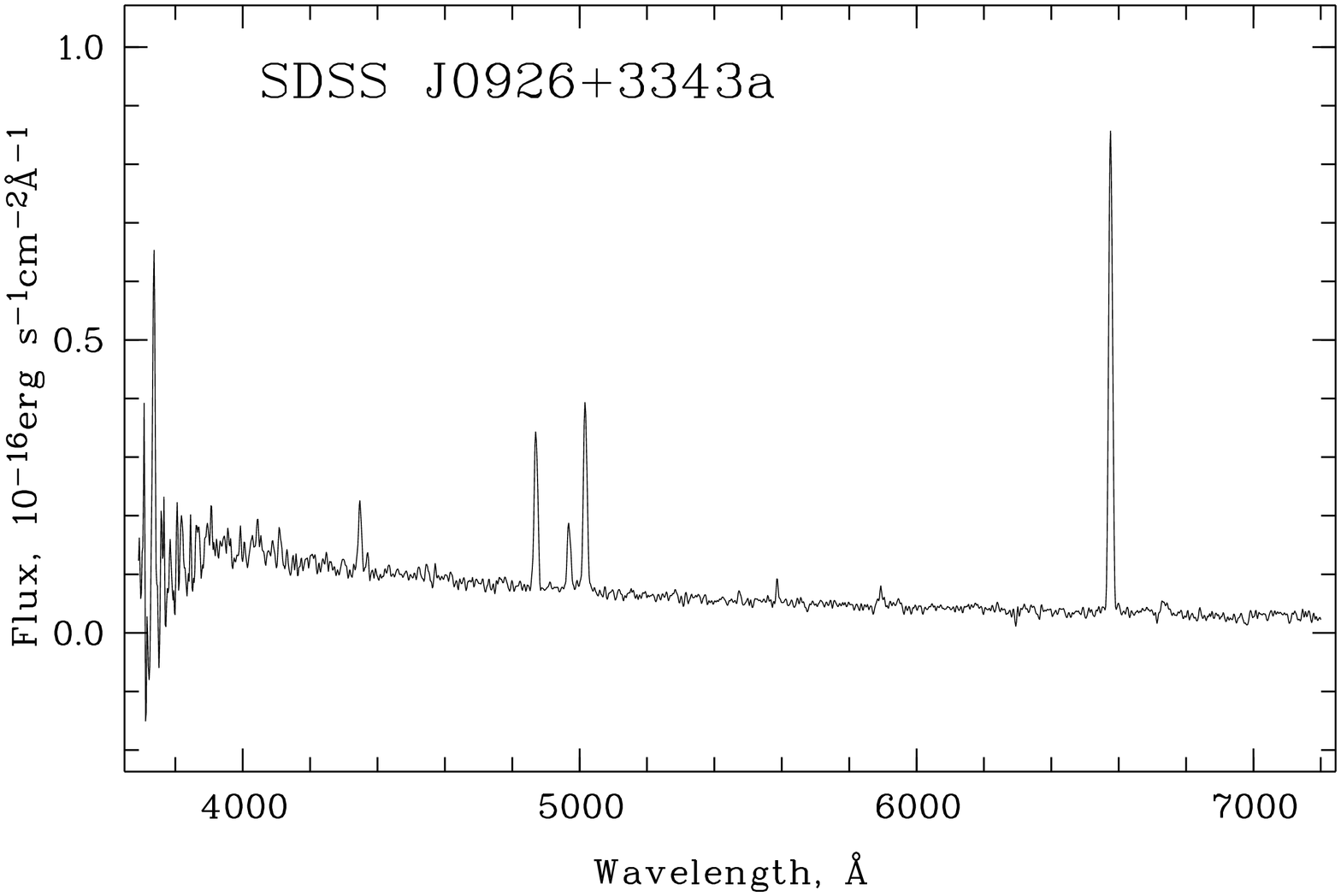}
 \includegraphics[angle=-90,width=7.5cm, clip=]{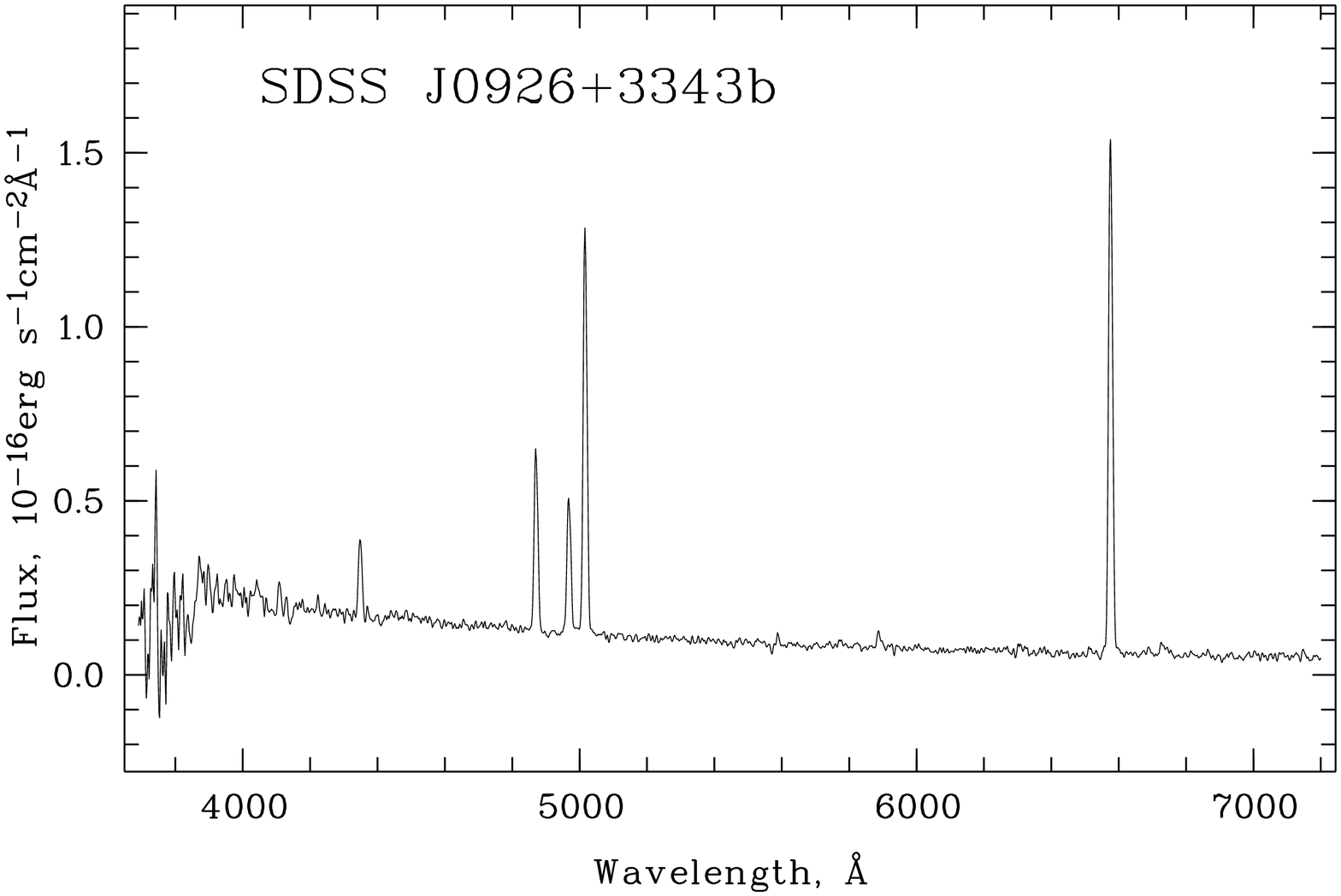}
  \caption{
Spectra with resolution of 12~\AA\  of regions {\bf a} (top panel) and
{\bf b} (bottom panel)
in SDSS J0926+3343, obtained on 2009 January 22.
}
	\label{fig:spectra}
 \end{figure}

The spectra of regions {\bf a} and {\bf b}, obtained on 2009.01.22, are shown
in the top and bottom panels of Fig.~\ref{fig:spectra}, respectively.
The fluxes of these two regions are low, with a S-to-N ratio in
the principal line [\ion{O}{iii}] $\lambda$4363
of $\sim$2--3 only. Due to the strong noise in the ultra-violet, the
important line [\ion{O}{ii}] $\lambda$3727
has also a low S-to-N ratio. The situation is similar for
spectra obtained during the other nights.

The main goal of our spectral observations was to get the most reliable
estimate of the galaxy parameter O/H from the data on the \ion{H}{ii} regions
{\bf a} and {\bf b}. Due to the low quality of individual measurements,
it is necessary to average all of them together. However, the seeing during
two of four nights was somewhat poor (1\farcs9 - 2\farcs3)
to resolve these regions as individual in our 2D spectra.
To use all spectral data in the most efficient way, we employ the following
approach.

First, we have checked how robust is the estimate of O/H for
two adjacent \ion{H}{ii} regions with the same O/H, but somewhat different
physical conditions and emission line relative intensities, if we measure
them first separately and second, as one entity. We have simulated
the artificial `sum' spectrum of this one entity, just summing up fluxes
of all lines in the two individual spectra. The latter were chosen to have
the line intensity ratios similar to the observed ones. As one could expect
from the general principle of analytic functions, the resulting physical
parameters
appeared half-way between those for individual objects, and the O/H, in turn,
to be very close (within 0.01~dex) to the single value of the two
individual objects.
Since the sum spectra of regions {\bf a} and {\bf b} also have better S-to-N
ratio than the spectra of individual components, assuming that their O/H are
the same, we decided to extract in all four independent 2D spectra the same
region of 21 rows (or 7.5\arcsec) along the slit, which includes both regions
and gives us 1D spectra with the sum of fluxes of both regions.

In Table~\ref{tab:Intens1}, we present the line intensities F($\lambda$)
of all relevant emission lines measured in all four obtained `composite'
spectra, integrated over the region with extent of 21 pixel along the slit,
which includes both regions {\bf a} and {\bf b}, normalised by the intensity
of H$\beta$, and I($\lambda$), corrected for the foreground extinction
C(H$\beta$) and the equivalent widths of underlying Balmer absorption lines
EW(abs).

In Table~\ref{t:Chem1} we present  the electron temperatures for
zones of emission of [\ion{O}{iii}] and  [\ion{O}{ii}], the accepted electron
densities N$_{\rm e}$, and the ionic abundances of oxygen, along with the
total abundances derived for the above measured line intensities with the
classic \Te\ method, according to the scheme described in \citet{Kniazev08}.

To emphasise the consistency of the results of all nights, we summarise the
values of O/H for these four `composite' spectra with some relevant
spectral parameters in the top part of Table~\ref{tab:composit}.
The values of O/H are the same within their quite large uncertainties,
with the range of
12+$\log$(O/H) of 7.08\p0.19 to 7.25\p0.19. We accept their weighted mean
as the best estimate of the galaxy O/H general mean, which corresponds to
12+$\log$(O/H)=7.16\p0.05.

The S-to-N ratio in the principal faint line [\ion{O}{iii}] $\lambda$4363
is quite low in all analysed spectra; this leads to large errors in \Te\ and,
respectively, to large errors in the derived O/H values. In such cases, or
when [\ion{O}{iii}] $\lambda$4363 is not detected,  in the low-metallicity
regime,
\citet{IT07} proposed the so-called semi-empirical method which employs
the relation between the total relative intensity of lines
[\ion{O}{iii}] $\lambda\lambda$4959,5007 and [\ion{O}{ii}] $\lambda$3727 and
\Te\
\citep[see the similar method for the high-metallicity regime in][]
{Pagel79,Shaver83}.
This method was tested on several of the most metal-poor
\ion{H}{ii}-regions in \citet{IT07} and also by us on our own data.
The O/H values, derived by this method, appeared to be well consistent, within
rather small errors, with those derived via the direct T$_{\rm e}$-method.
The calculations of O/H are performed on the same formulae as described
above in the direct method, but instead of \Te\ derived through the
intensity ratio of [\ion{O}{iii}] lines $\lambda$4363 and
$\lambda\lambda$4959,5007,
we accept \Te\ derived with the empirical formula from \citet{IT07}.
The resulting values of O/H for each of the nights are shown in the bottom of
Table~\ref{tab:composit}. The respective values of 12+$\log$(O/H) vary in
a significantly narrower range of 7.08\p0.10 to 7.12\p0.05. Their
weighted mean corresponds to 12+$\log$(O/H)=7.11\p0.01.
The latter estimate is partly independent on that derived with the
direct T$_{\rm e}$-method since it does not use the intensity of
[\ion{O}{iii}] $\lambda$4363 line. Therefore, both the estimates of the
galaxy O/H
can be further averaged (with weights), that results in the `final' value of
O/H in SDSS~J0926+3343 of 12+$\log$(O/H)=7.12$\pm$0.02.

\begin{table*}
\centering
\caption{Spectral parameters and O/H for the four SDSS~J0926+3343 `composite'
spectra}
\label{tab:composit}
\begin{tabular}{lcccc} \\ \hline
Run/Date       &2008.11.26 &2009.01.21 &2009.01.22 &2009.02.19   \\ \hline
\multicolumn{5}{p{9.0cm}}{\bf O/H derived via direct T$_{\rm e}$ method}  \\  \hline
C(H$\beta$)    & 0.00\p0.06 & 0.00\p0.07 & 0.14\p0.04 & 0.10\p0.04   \\
EW(abs)        & 2.8\p0.7  & 0.0\p2.4  & 5.4\p0.4  & 5.9\p0.4     \\
I(4363)/I(\Hb)  &0.064\p0.029&0.060\p0.027&0.046\p0.017&0.052\p0.017  \\
I(4959+5007)/I(\Hb) & 2.28\p0.09 & 2.44\p0.09 & 2.52\p0.07 & 2.40\p0.06  \\
I(3727)/I(\Hb)  & 0.55\p0.09 & 0.24\p0.29 & 0.43\p0.10 & 0.59\p0.08   \\
O/H (in 10$^{-5}$)&1.197\p0.515&1.273\p0.678&1.780\p0.769&1.515\p0.441  \\
12+$\log$(O/H) & 7.08\p0.19 & 7.10\p0.23 & 7.25\p0.19 & 7.18\p0.13  \\
\hline
\multicolumn{5}{p{9.0cm}}{\bf O/H derived via semi-empirical method of Izotov \& Thuan}  \\  \hline
O/H (in 10$^{-5}$)&1.331\p0.156&1.199\p0.278&1.285\p0.157&1.322\p0.149  \\
12+$\log$(O/H) & 7.12\p0.05 & 7.08\p0.10 & 7.11\p0.05 & 7.12\p0.05  \\
\hline
\hline
\end{tabular}
\end{table*}

\subsection[]{H{\sc i} parameters}

\begin{figure}
  \centering
 \includegraphics[angle=-90,width=7.0cm, clip=]{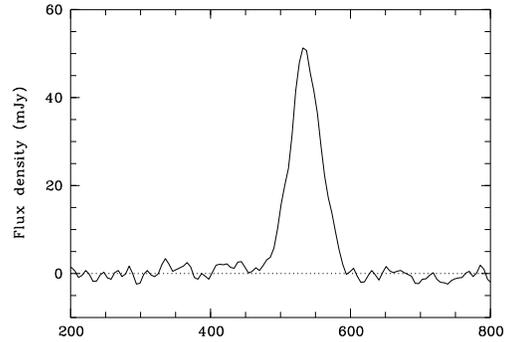}
  \caption{
The NRT profile of  H{\sc i}-line emission from  SDSS J0926+3343. X-axis
shows
radial heliocentric velocity in \kms. Y-axis shows the galaxy flux density
in mJy.
}
	\label{fig:HI}
 \end{figure}

The profile of the 21-cm H{\sc i} line emission in SDSS J0926+3343 obtained
with
the NRT is shown in Fig.~\ref{fig:HI}. Its parameters are as follows.
The integrated  H{\sc i} flux F(H{\sc i})=2.54$\pm$0.07 Jy~\kms.
The central velocity of the  profile is 536$\pm$2~\kms.
The profile widths are
W$_\mathrm{50}$=47.4$\pm$3~\kms\ and W$_\mathrm{20}$=80.5$\pm$7~\kms.

To estimate the galaxy global parameters, we accepted for its distance
the value $D$=10.7~Mpc (with the respective scale of 52 pc in 1\arcsec).
The latter comes from its V$_{\rm LG}$=488~\kms, the accepted Hubble constant
of 73~\kms~Mpc$^{-1}$ and the correction for the large negative peculiar
velocity, discussed by \citet{Tully08}, which we adopt for this region
as 290~\kms.
The  H{\sc i} mass of the galaxy is determined by the well-known relation for
optically thin  H{\sc i}-line emission from \citet{Roberts69}
that gives M(H{\sc i})=6.8$\times$10$^{7}$~M\sunn.

\subsection{Photometric properties and the age estimates}
\label{sect:age}

\begin{figure}
  \centering
 \includegraphics[angle=-90,width=7.5cm, clip=]{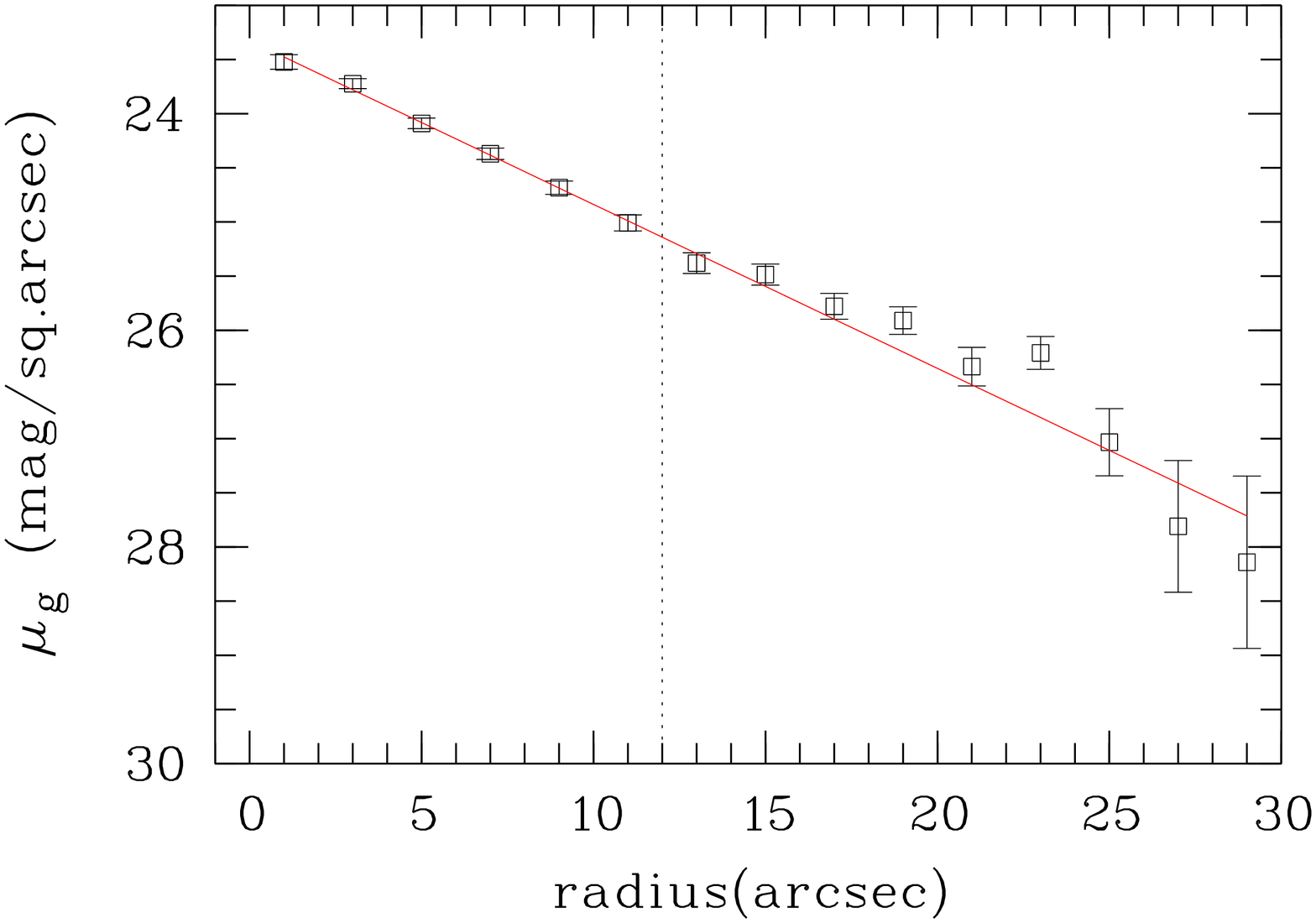}
 \includegraphics[angle=-90,width=7.5cm, clip=]{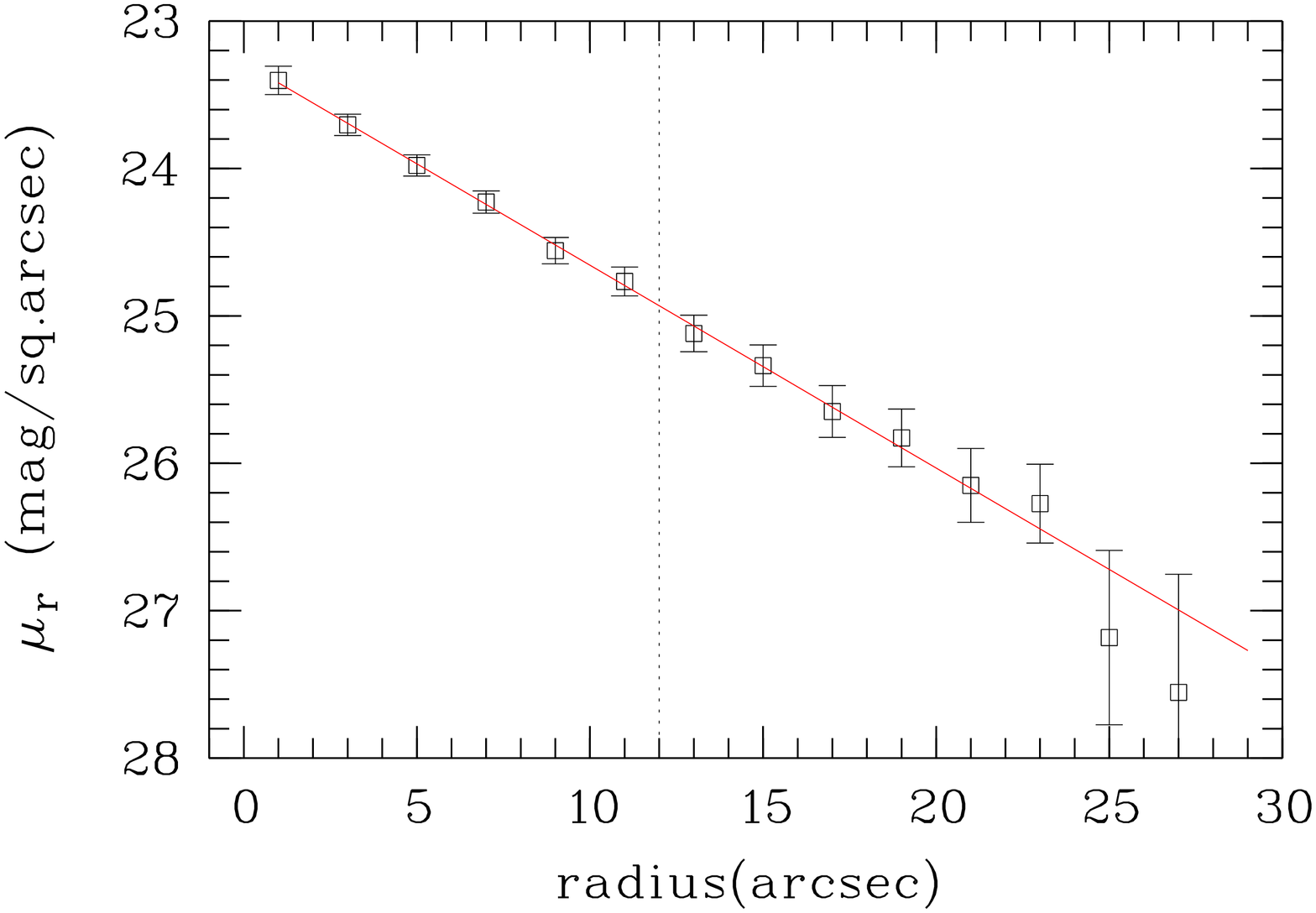}
 \includegraphics[angle=-90,width=7.5cm, clip=]{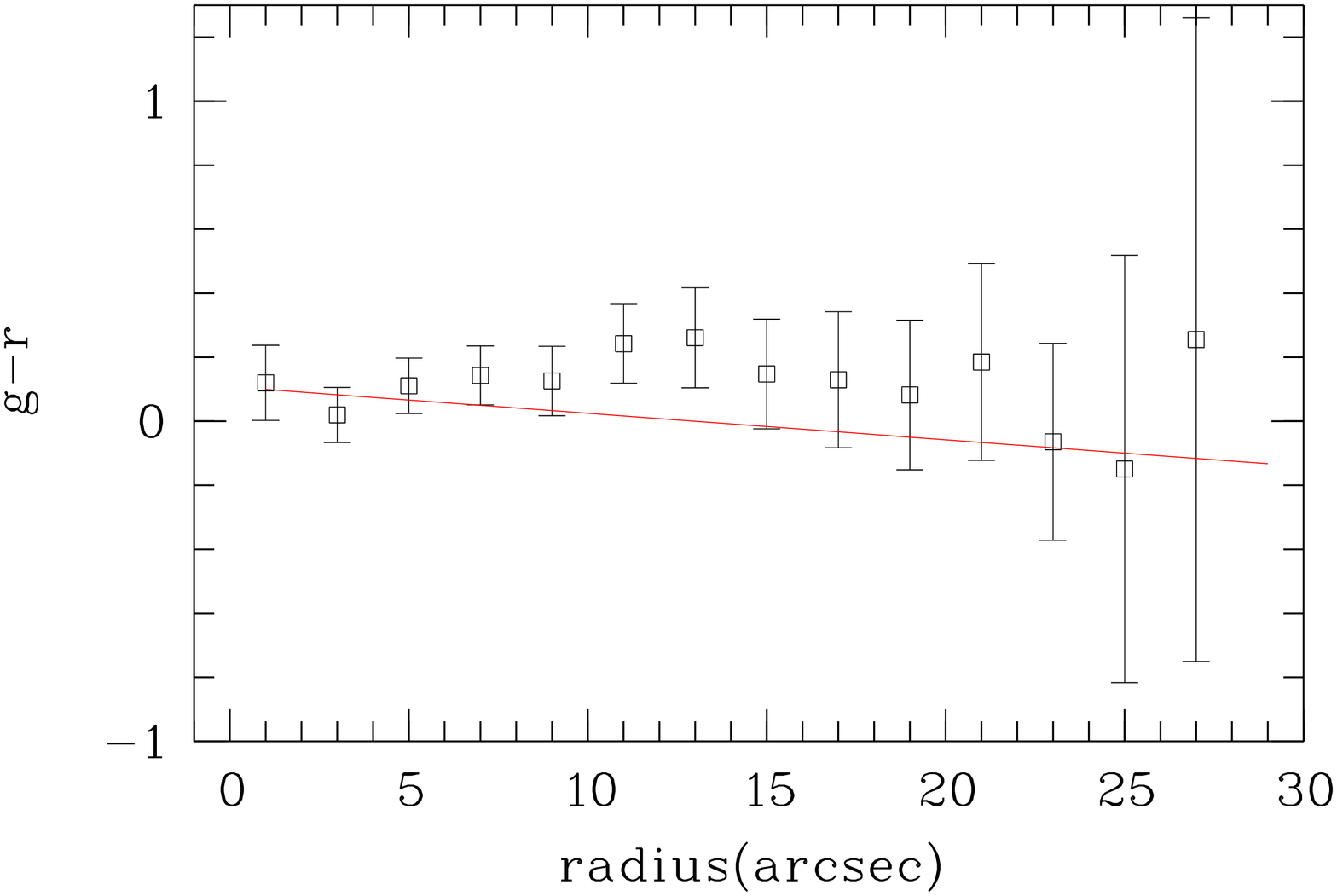}
  \caption{
SDSS J0926+3343 surface brightness in $g$ and $r$ filters and $g-r$ colour
versus the effective radius. The solid lines show model fits for exponential
discs on the profile part with $<$12\arcsec\ (shown by vertical dotted line).
}
	\label{fig:SB_prof}
 \end{figure}

From the surface brightness radial profiles in $g$ and $r$ filters shown
in Fig.~\ref{fig:SB_prof}, one can derive the colour of the underlying
`exponential disc' $(g-r)$=$\sim$0.10. Using transformation formula of
\citet{Lupton05}, this translates to the relation
$\mu_{\rm B}$=$\mu_{g}$+0.25~mag~arcsec$^{-2}$.
Then, from the galaxy SB profile in $g$-filter, the `optical' radius
R$_{\rm opt}$ at $\mu_{\rm B}$=25~mag~arcsec$^{-2}$
and the Holmberg radius R$_{\rm Ho}$ at $\mu_{\rm B}$=26.5~mag~arcsec$^{-2}$
are R$_{\rm opt,eff}$=9\farcs4  and R$_{\rm Ho,eff}$=19\farcs7.
To transform these effective radii to real ones, the correction factor
$(b/a)^{-1/2}$=1.89 should be applied, where the galaxy axial ratio
$b/a$=0.28 was adopted from NED/SDSS. This results in
the `optical' radius of R$_{\rm opt}$=17\farcs8 ($\sim$0.93~kpc)
and in the `Holmberg' radius of R$_{\rm Ho}$=37\farcs2 ($\sim$1.94~kpc).

From the above exponential disc fits (for internal regions with
$R <$12\arcsec) we also estimated the central SB and disc scale lengths.
The best fits are for $g$ and $r$ filters, which give $\mu_{\rm g}^0$=23.32
and
$\mu_{\rm r}^0$=23.28~mag~arcsec$^{-2}$ with respective scale lengths of
6\farcs6 and 7\farcs2. Their mean 6\farcs9 corresponds to the linear
scale length of 0.36 kpc. When corrected for $(b/a)^{-1/2}$, this gives a
radial disc scale length of 0.68 kpc.
With the same magnitude transform as above, we derive the {\it observed}
central blue SB $\mu_{\rm B}^0$=23.48~mag~arcsec$^{-2}$ (after correction
for A$_{\rm B}$=0.08). Due to almost edge-on orientation of the galaxy disc,
its observed central SB is significantly enhanced. We correct the latter
value, taking the visible axial ratio $p=b/a$=0.28 and adopting the internal
axial ratio $q=$0.20. The inclination angle $i$, derived from the well known
formula: $\cos(i)^{2} = (p^{2}-q^{2})/(1-q^{2})$, amounts to $i=$78.5\degr.
The respective correction for $\mu_{\rm B}^0$ is equal to
1.75~mag~arcsec$^{-2}$. Thus, the central SB corrected for inclination
appears very low: $\mu_{\rm B}^0$=25.39~mag~arcsec$^{-2}$, and this galaxy
should be classified as a `Very Low Surface Brightness'.

The measured $u,g,r,i$ magnitudes and the derived, Galaxy extinction corrected
colours $(u-g), (g-r)$ and $(r-i)$ are presented in Table~\ref{tab:photo}.
In the first line we give the parameters for the whole galaxy. In the second
line -- for the whole galaxy with removal of the light of \ion{H}{ii}
regions {\bf a, b, c} and the small blue nebulosity {\bf d} somewhat outside
the main disc, at $\sim$5\arcsec\  west of region {\bf b}. The parameters in
the two last lines are discussed below.

Our main goal is to compare the observed colours of stellar population with
the PEGASE2 model evolutionary tracks \citep{pegase2}, in order to obtain
the estimates of the
maximal stellar ages in the galaxy. As one can see from the $(g-r)$ colour
radial profile in Fig.~\ref{fig:SB_prof},  due to the contribution of the
mentioned \ion{H}{ii} regions to the light of the outer parts of the galaxy,
it is difficult to estimate from the profile the outer colours and decide
whether there is a colour gradient in the underlying stellar population.
Also, due to the very elongated form of the galaxy, the photometry in
ring diaphragms mixes the light from different parts of the `disc', and thus
will tend to wash-up colour gradients if they are present. Therefore, to
analyse the stellar light colours of the galaxy outer parts, we employ the
approach, used in our analysis of the stellar colours in the galaxy DDO~68
\citep{DDO68_sdss}.

First, to address the possible colour differences, we measured the light of
the central `ridge', the elongated structure along the `disc' major axis with
somewhat enhanced SB. This was done by summing up the light of ten round
diaphragms with 3 pixels radius ($\sim$1\farcs2) placed along the
`ridge'.  Its parameters are given in the third line of Table~\ref{tab:photo}.
As one can see, the `ridge' colours are very close to those of the total
`disc', with the $(g-r)$ colour probably to be a bit bluer. To estimate the
colours of the `outer' region, we performed the similar photometry in the
regions, adjacent to the `ridge', summing-up the light from twelve diaphragms
of the same size, in which the signal in all filters was above the background
noise level. This area was limited by the extent in $u$-filter.
The derived parameters of this `outer' region are shown in the fourth line
of Table~\ref{tab:photo}.
Its $(u-g)$ colour appears marginally redder, while the $(g-r)$ and $(r-i)$
are consistent within their errors with the colours of the `ridge' and the
whole galaxy.

\begin{table*}
\centering
\caption{Magnitudes and colours of SDSS J0926+3343 and its regions}
\label{tab:photo}
\begin{tabular}{lccccccc} \\ \hline
Region          &$u$\p $err_u$&$g$\p $err_g$&$r$\p $err_r$& $i$\p $err_i$ &$(u-g)^0$ \p$err$&$(g-r)^0$ \p$err$ &$(r-i)^0$ \p$err$ \\ \hline
Total light     & 17.95 0.08  & 17.08 0.03  &16.98 0.04   & 16.96 0.04    &   +0.87 0.08    & +0.08 0.05       & -0.06 0.06       \\   
Total with removed \ion{H}{ii}& 18.04 0.08&17.14 0.03&17.03 0.04& 17.01 0.04& +0.88 0.08    & +0.09 0.05       & +0.01 0.06       \\
The `ridge'     & 20.44 0.11  & 19.54 0.04  &19.53 0.06   & 19.55 0.08    &   +0.88 0.11    & -0.01 0.07       & -0.03 0.10       \\
The outer region& 20.55 0.14  & 20.00 0.06  &19.90 0.10   & 19.90 0.13    &   +0.52 0.15    & +0.08 0.12       & -0.02 0.16       \\
\hline
\end{tabular}
\end{table*}

\begin{figure}
  \centering
 \includegraphics[angle=-90,width=9.0cm, clip=]{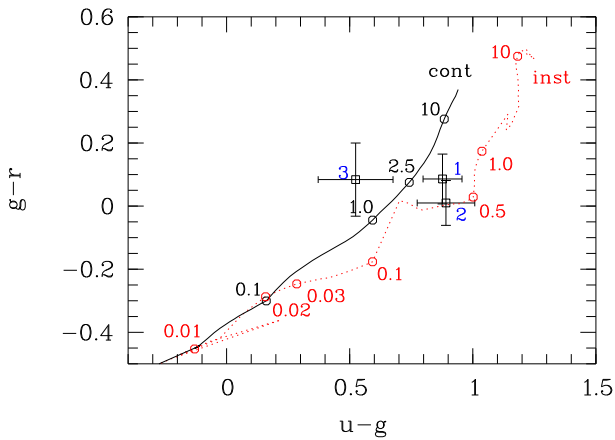}
  \caption{
The $(u-g), (g-r)$ colours of SDSS J0926+3343, corrected for the small
extinction in the Galaxy, compared with the PEGASE2 evolutionary tracks
for the standard Salpeter IMF and Z=0.0004. The two extremes of the SF law
are shown: continuous with constant SFR (solid line) and instantaneous SF
(dashed line). Open circles along the tracks mark the time in Gyr elapsed
since the beginning of SF. The point labelled ``1'' indicates the integrated
colours of the galaxy, excluding the four mentioned small \ion{H}{ii}-regions.
The point labelled ``2'' represents the colours of the `ridge'
(Fig.~\ref{fig:image}, top panel). The point labelled ``3'' represents the
colours of the outer part adjacent to the `ridge' region (see description
in text). If interpreted in terms of the colours, nearest to a model
evolutionary track, the latter best correspond to the continuous SF for
ages of T$\sim$1--3~Gyr.
}
	\label{fig:photo}
 \end{figure}

In Fig.~\ref{fig:photo} we compare the derived colours in
Table~\ref{tab:photo} with the model tracks from the PEGASE2 package. We used
the standard Salpeter IMF and the metallicity Z=0.0004, which is the best
proxy to the value found from our spectroscopy of \ion{H}{ii} regions.
We show the
evolutionary tracks for the two extreme SF laws to derive the upper limit of
the visible stellar population ages. The integrated colours of the disc `sit'
in between of continuous and instantaneous SF law tracks. Its rather blue
colour $(g-r)$ corresponds to ages from $\sim$0.7 to $\sim$2.5 Gyr, depending
on the SF law. The $ugr$ colours of the `ridge' are better consistent with
instantaneous starburst occurred several hundred Myr ago. The $ugr$ colours
of the `outer' region appear a bit offset from the continuous SF track, but
accounting for the observed colour errors, the difference is not significant.
The nearest part of the track (in terms of amount of $\sigma_{\rm err}$ and
the related probability) corresponds to the ages of stellar population of
$\sim$1--3~Gyr. That is, within the limits of measurable LSB periphery,
the oldest visible stellar population appears quite young ($\sim$2~Gyr),
in difference with the absolute majority of thousands galaxies, for which
the similar data are acquired. In the $gri$ colour plot (not shown) both
tracks go too close each to other, so the galaxy $gri$ colours alone would
be inconclusive. However, the measured $(r-i)$ colours are consistent with
the above conclusions derived from $ugr$ colours.

\section[]{DISCUSSION}
\label{sec:dis}

In Table~\ref{tab:param} we present the main parameters of the studied
galaxy. From the total magnitudes in filters $g$ and $r$ (Table
\ref{tab:photo}), with the transformation equations of \citet{Lupton05},
we derive the total $B$-band magnitude, $B_{\rm tot}$=17.34$\pm$0.03.
For the accepted distance modulus of SDSS J0926+3343 $\mu$=30.16 (D=10.7 Mpc
and A$_{\rm B}$=0.08) its absolute magnitude M$_{\rm B}^0 = -$12.90.
The latter corresponds to L$_{\rm B}$=2.24$\times$10$^{7}$~L$_{\rm B}$\sunn.
Then, from the M(H{\sc i}) derived in the previous section, one obtains the
ratio M(H{\sc i})/L$_{\rm B} \sim$3.0 (in solar units). This value is rather
large
for the Local Universe galaxies and is among the few per cent highest ratios
ever measured.

From the H{\sc i} line width at 20\%-level of the maximum intensity,
W$_\mathrm{20}$, one can estimate the maximal rotational velocity, using the
standard approximation, as, e.g., formula (12) from \citet{Tully85}. Since the
galaxy is seen almost edge-on, the inclination correction is only 1.02.
For a measured line width W$_\mathrm{20}$=80.5~\kms, the derived
V$_{\rm rot}$
$\sim$31.2~\kms, and the inclination corrected V$_{\rm rot} \sim$32~\kms,
which is typical of dwarf galaxies with comparable
M$_{\rm B}$  in the faint dwarf sample FIGGS of \citet{Begum08b}.
Having V$_{\rm rot}$ and the characteristic size of the galaxy, one can
estimate its total (dynamical) mass which is necessary to balance the
centrifugal force within a certain radius.

The typical radii of H{\sc i} discs (at the column  density level of
10$^{19}$~atoms~cm$^{-2}$) in dwarf galaxies with M$_{\rm B} \sim$-13,
close to that of this LSB dwarf, are 2.5-3 times larger than the Holmberg
radius \citep[e.g.,][]{Begum08a,Begum08b}. Therefore, we accept that the
H{\sc i} radius of this galaxy is 2.7~R$_{\rm Ho}$, that is
$R_{\rm HI} =$5.2~kpc. Then, from the relation
M($R < R_{\rm HI}$) = V$_{\rm rot}^2~\times$~$R_{\rm HI}$/$G$, where $G$ is
the gravitational constant, one derives the total mass within
$R_{\rm HI}$ as follows:
M$_{\rm tot}$=12.4$\times$10$^{8}$~M\sunn.
%
To get the estimate of the galaxy gas mass, we sum M(H{\sc i}) and M(He) (a
fraction of 0.33 of H{\sc i} mass) that gives
M$_{\rm gas}$=9.04$\times$10$^{7}$~M\sunn.

The mass of stars can be estimated as follows. We take from
Table~\ref{tab:photo} the total $g$-magnitude of the LSB disc (without light
of \ion{H}{ii} regions), equal to 17.14. We also account for the light
contribution of the younger stellar population of the `ridge', of the order
of 10~\% in $g$-filter. Then, the improved estimate of the LSB disc
$g$-filter magnitude is 17.24. This corresponds to the extinction-corrected
absolute magnitude M$_{\rm g}$=--13.0.
According to the PEGASE2 model track with continuous SF and constant SF rate,
for T=2.5~Gyr, the colour $g-V$=--0.08, that is its M$_{\rm V}^0$=--12.92.
Accounting that for this PEGASE2 track, the luminosity of a star cluster
with T=2.5~Gyr corresponds to the luminosity of $V$=6.47 mag per 1~M\sunn,
which is fainter than the galaxy estimated M$_{\rm V}^0$ by 19.39 mag,
we obtain that the full stellar mass (with T=2.5~Gyr) of the galaxy is
M$_{\rm star} \sim$5.70$\times$10$^{7}$~M\sunn.
This implies that the total baryonic mass
M$_{\rm bary}$=1.47$\times$10$^{8}$~M\sunn\ and the  gas mass-fraction
$\mu_{\rm g}$=M$_{\rm gas}$/(M$_{\rm gas}$+M$_{\rm star}$)=0.61.
With the mass ratio M$_{\rm tot}$/M$_{\rm bary}$=8.3, this LSBD galaxy is
clearly a Dark-Matter dominated one. To get a better precision estimate of
the DM halo mass and its radial distribution, one needs H{\sc i} mapping
of the galaxy.

Having all these unusual properties  in one object (the record-low value of
O/H, `young' (T$\sim$2$\pm$1~Gyr) main stellar population, large gas-mass
fraction), the galaxy SDSS J0926+3343 represents a nice nearby laboratory
to study several cosmologically related issues. This looks to be very
isolated dwarf galaxy (the nearest luminous galaxies are at D $>$ 2 Mpc)
and, thus, provides a good opportunity to study the SF, induced presumably
by the internal instabilities in the most metal-poor environment and at
quite low disc baryonic surface densities. The star `clusters'
or `stellar associations' responsible for ionisation of \ion{H}{ii} regions
{\bf a} and {\bf b}, according to the values of their EW(H$\beta$) are quite
young, having  ages of 7--8 Myr.
Therefore, one can hope to find in them individual
massive stars with Z $\sim$Z\sunn/35, similar to, e.g., the LBV star in that
low-Z dwarf galaxy DDO~68 \citep{Pustilnik_LBV,Izotov_LBV}.

Studying gas kinematics in this LSBD may determine the density distribution
of the DM halo, and thus, better constrain the DM nature. There is still
a chance that this LSBD galaxy experiences a substantial external
disturbance, since its disc looks somewhat warped, and
its main SF sites are situated at large distances from the galaxy centre.
However, the alternative option of triaxial DM halo can also
explain the visible warps in this LSBD disc \citep[e.g.,][]{jeon09}.
In particular, H{\sc i} maps with the angular resolution of 5\arcsec\ -
10\arcsec\
would help to elucidate the neutral gas morphology and kinematics and to
check the hypothesis of an intergalactic H{\sc i} cloud infall or a minor
merger.

Looking at this unusual  galaxy in a more general aspect, one can notice
its appearance in the nearby Lynx-Cancer void, where other very metal-poor
and unevolved dwarf galaxies are found. They include, in particular,
DDO~68 at the mutual distance to SDSS J0926+3343 of only $\sim$1.6~Mpc
\citep{DDO68}, and HS~0822+3542 and SAO~0822+3542 \citep{SAO0822}.
A few XMD dwarf galaxies in the region of this void found by \citet{IT07},
as well as several more such galaxies found by us, will be presented in a
forthcoming paper.

In Table \ref{tab:param}, along with the data for SDSS J0926+3343, we present
the properties of DDO~68, in order to emphasise the range of parameters
in two `neighbouring' most metal-poor dwarf galaxies residing in the
Lynx-Cancer void.
The blue luminosity and the gas mass of DDO~68 are higher by an order of
magnitude, while its optical size is larger by a factor of $\sim$5. The
DDO~68 morphology, kinematics and stellar colours imply that its appearance
is related to the recent merger ($\sim$1 Gyr ago) of two very gas-rich and
very metal-poor dwarf galaxies (or gas clouds/protogalaxies).

The increased density of such rare objects in this region is certainly not
by chance. XMD galaxies seem to prefer residence in  some selected regions.
Our findings indicate that while the fraction of such galaxies is small,
namely the void environment is favourable for the retarded dwarf galaxy
formation and their slower evolution. The well-known
interacting/merging pair of the most metal-poor dwarf galaxies
SBS~0335--052E,W
\citep[see new data in][]{Ekta09,Izotov09} is also situated near the border
of a large void \citep{Peebles01}. The existence of the sizable H{\sc i} cloud
population
in `nearby' voids, with the baryonic masses comparable to those of dwarf
galaxies,
visible through their Ly-$\alpha$ absorption \citep[][ and references therein]
{Manning02,Manning03}, also hints on the unevolved state of the significant
fraction of baryons in voids. The more advanced analysis of the Lynx-Cancer
void dwarf galaxy census and the summary of their properties will be
presented elsewhere.

\begin{table}
\caption{Main parameters of SDSS J0926+3343 and DDO~68}
\label{tab:param}
\begin{tabular}{lcc} \\ \hline
Parameter                           &  J0926+3343             &   DDO~68         \\ \hline
R.A.(J2000.0)                       & 09 26 09.45             & 09 56 45.7       \\
DEC.(J2000.0)                       & $+$33 43 04.1           & $+$28 49 35     \\
A$_{\rm B}$ (from NED)              & 0.08                    & 0.08             \\
B$_{\rm tot}$                       & 17.34$\pm$0.03$^{(2)}$  & 14.74            \\
V$_{\rm hel}$(H{\sc i})(\kms)       & 536$\pm$2$^{(3)}$       & 502$\pm$2$^{(9)}$  \\
V$_{\rm LG}$(H{\sc i})(\kms)        & 488$\pm$2$^{(3)}$       & 428              \\
Distance (Mpc)                      & 10.7$^{(2)}$            & 9.9$^{(2)}$      \\
M$_{\rm B}^0$ $^{(4)}$              &  --12.90                & --15.32$^{(2)}$  \\
Opt. size (\arcsec)$^{5}$           & 35.8$\times$9.9$^{(2)}$ & 103$\times$38$^{(8)}$ \\
Opt. size (kpc)                     & 0.93$\times$0.26$^{(2)}$& 4.94$\times$1.82  \\
$\mu_{\rm B}^0$(mag~arcsec$^{-2}$)  & 25.4$^{(2)}$            & $<$22.7$^{(8)}$   \\
12+$\log$(O/H)                      & 7.12$\pm$0.02$^{(2)}$   & 7.14$\pm$0.03$^{(10)}$  \\
H{\sc i} int.flux$^{(6)}$           & 2.54$\pm$0.07$^{(2,3)}$ & 28.9$\pm$3.0    \\
W$_\mathrm{50}$ (km s$^{-1}$)       & 47.4$\pm$3$^{(2,3)}$    & 82$^{(9)}$         \\
W$_\mathrm{20}$ (km s$^{-1}$)       & 80.5$\pm$7$^{(2,3)}$    & 105$^{(9)}$        \\
V$_\mathrm{rot}$ (H{\sc i})(\kms)   & 32$^{(2,3)}$            & 52$^{(9)}$                \\
M(H{\sc i}) (10$^{7} M_{\odot}$)    & 6.8$^{(2)}$             & 66.8$^{(2)}$     \\
M$_{\rm dyn}$ (10$^{7} M_{\odot}$)  & 124$^{(2)}$             & 487              \\
M(H{\sc i})/L$_{\rm B}$$^{(7)}$     & 3.0$^{(2)}$             & 2.9$^{(9)}$      \\
T(main star population)             & 1--3~Gyr$^{(2)}$        & $\lesssim$1~Gyr$^{(11)}$      \\
\hline
\multicolumn{3}{p{8.2cm}}{%
(1) -- from NED; (2) -- derived in this paper;
(3) -- derived from NRT HI profile;  (4) -- corrected for Galactic extinction
A$_{\rm B}$=0.08; (5) -- $a \times b$ at $\mu_{\rm B}=$25\fm0~arcsec$^{-2}$;
(6) -- in units of Jy$\cdot$\kms; (7) -- in solar units; (8) \citet{DDO68};
(9) - \citet{Ekta08}; (10) - \citet{IT07};
(11) - \citet{DDO68_sdss}. }
\end{tabular}
\end{table}

Summarising the results and discussion above, we draw the following
conclusions:

\begin{enumerate}
\item
The edge-on VLSB ($\mu_{\rm B}^0$=25.4~mag~arcsec$^{-2}$)
dwarf galaxy SDSS J0926+3343 is situated in the
nearby Lynx-Cancer void. It has several faint \ion{H}{ii} regions.
For the two brightest of them, {\bf a} and {\bf b}, at the NE edge, we
obtained optical spectra and derived their oxygen abundance.
The general means of their O/H values derived by the direct T$_{\rm e}$ and
by the semi-empirical method of \citet{IT07} are well consistent
with each other, yielding 12+$\log$(O/H) of 7.16\p0.05 and 7.11\p0.01,
respectively.
We accepted the mean of the two,  12+$\log$(O/H)=7.12$\pm$0.02, as
the measure of O/H in the galaxy.
\item
SDSS J0926+3343 appeared the most metal-poor galaxy in the Local Universe,
situated in the remarkable proximity ($\sim$1.6~Mpc) to another that low
metallicity dwarf galaxy DDO~68 (with 12+$\log$(O/H)=7.14).
\item
The HI integrated flux and the velocity width data for this galaxy, along with
the optical photometry and the size estimates, indicate the very gas-rich
object, with M(HI)/L$_{\rm B} \sim$3.0 and with the derived gas mass-fraction
of $\mu_{\rm g} \sim$0.6. The galaxy is a DM dominated, with the estimated
mass ratio of M$_{\rm tot}$/M$_{\rm bary} \sim$8.3.
\item
The $(u-g)$, $(g-r)$ and $(r-i)$ colours of SDSS J0926+3343 are `blue'. The
`outer' region colours well match the PEGASE2 model track for the evolving
stellar population with continuous SF for ages of T$\sim$1--3 Gyr. Thus,
all three observational parameters: O/H, $\mu_{\rm g}$ and blue colours
are consistent with an evolutionary young status of this LSB dwarf galaxy.
\end{enumerate}

\section*{Acknowledgements}

SAP and ALT acknowledge the support of this work through the RFBR grant
No. 06-02-16617. SAP is also grateful for the support through the Russian
Federal
Agency of Education grant No. 2.1.1/1937. SAP, ALT and AYK acknowledge the
BTA TAC for the continuous support of this project at the SAO 6-m telescope.
The authors thank A.~Valeev for the help with BTA observation. SAP and JMM
acknowledge the NRT TAC for allocation of time for this program in 2007.
The authors thank the reviewer D. Saikia for useful suggestions which
improved the paper presentation.
The authors acknowledge the spectral and photometric data and the related
information available in the SDSS database used for this study.
The Sloan Digital Sky Survey (SDSS) is a joint project of the University of
Chicago, Fermilab, the Institute for Advanced Study, the Japan Participation
Group, the Johns Hopkins University, the Max-Planck-Institute for Astronomy
(MPIA), the Max-Planck-Institute for Astrophysics (MPA), New Mexico State
University, Princeton University, the United States Naval Observatory, and
the University of Washington. Apache Point Observatory, site of the SDSS
telescopes, is operated by the Astrophysical Research Consortium (ARC).
This research has made use of the NASA/IPAC Extragalactic
Database (NED), which is operated by the Jet Propulsion Laboratory,
California Institute of Technology, under contract with the National
Aeronautics and Space Administration.


\bsp

\label{lastpage}

\end{document}